\documentclass[twocolumn,showpacs,preprintnumbers,amsmath,amssymb]{revtex4}

\usepackage{graphicx}
\usepackage{dcolumn}
\usepackage{bm}
\usepackage{graphicx}
\usepackage{amsmath}
\usepackage{amssymb}
\usepackage{datetime}

\begin{document}

\newcommand{\be}{\begin{equation}}
\newcommand{\ee}{\end{equation}}
\newcommand{\ac}{\emph{ac}-plane}
\newcommand{\ba}{\emph{b}-axis}
\newcommand{\sg}{\emph{R}$\overline{3}$}
\newcommand{\kvec}{{\bfseries{\emph{k}}}=0 }
\newcommand{\mub}{$\mu_{\mathrm B}$}
\newcommand{\TN}{$T_\mathrm{N}$}
\newcommand{\scbo}{SrCu$_{2}$(BO$_{3}$)$_{2}$}
\newcommand{\ypp}{Yb$_{2}$Pt$_{2}$Pb}
\title{Spin Liquids and Antiferromagnetic Order in the Shastry-Sutherland-Lattice Yb$_{2}$Pt$_{2}$Pb}

\author{M. S. Kim}
\affiliation{Department of Physics and Astronomy, Stony Brook University, Stony Brook NY 11794, USA}
\author{M. C. Aronson}
\affiliation{Brookhaven National Laboratory, Upton, New York 11973, USA}
\affiliation{Department of Physics and Astronomy, Stony Brook University, Stony Brook NY 11794, USA}

\begin{abstract}
We present measurements of the magnetic susceptibility $\chi$ and the magnetization M  of single crystals of metallic \ypp, where localized Yb moments lie on the geometrically frustrated Shastry-Sutherland Lattice (SSL). Strong magnetic frustration is found in this quasi-two dimensional system, which orders antiferromagnetically (AF) at T$_{N}$=2.02 \emph{K} from a paramagnetic liquid of Yb-dimers, having a gap $\Delta$=4.6 \emph{K} between the singlet ground state and the triplet excited states. Magnetic fields suppress the AF order, which vanishes at a 1.25 \emph{T} quantum critical point. The spin gap $\Delta$ persists to 1.5 \emph{T}, indicating that the AF  degenerates into a liquid of dimer triplets at T=0.  Quantized steps are observed in M(B) within the AF state, a signature of SSL systems. Our results show that \ypp~ is unique, both as a metallic SSL system that is close to an AF quantum critical point, and as a heavy fermion compound where geometrical frustration plays a decisive role.
\end{abstract}
 \maketitle

Much interest has focused on systems with geometrical frustration, where conventional antiferromagnetic (AF) order is suppressed in favor of more exotic ground states. The Shastry Sutherland Lattice (SSL) is one of the simplest frustrated systems~\cite{shastry1983}, consisting of planes of orthogonal dimers of moments with interdimer coupling $J^{\prime}$ and the intradimer coupling \emph{J}. The T=0 phase diagram has two limiting behaviors, depending on $J^{\prime}/J$. Nonordering dimers are found for small $J^{\prime}/J$, distinguished by an energy gap $\Delta$ between the singlet and triplet states of the dimer. Insulating \scbo~ exemplifies this disordered `spin liquid'(SL) regime~\cite{kageyama1999,miyahara2003,onizuka2000}. Conversely, AF order with gapless magnetic excitations is favored for large $J^{\prime}/J$, and the RB$_{4}$ (R= Gd,Tb,Dy,Ho,Er) compounds may represent this limit~\cite{etorneau1979,michimura2006,matas2010,iga2007}. A T=0 transition between the SL and AF phases has been predicted for \mbox{$J^{\prime}/J \simeq$0.6 - 0.7}~\cite{weihong1999,miyahara2003,alhajj2005,isacsson2006}, although symmetry-based arguments~\cite{carpentier2002} suggest that an intermediate state is required, such as a helical magnet~\cite{albrecht1996}, a weak SDW~\cite{carpentier2002}, or a plaquet ordered solid~\cite{koga2003}. The known SSL systems have so far not provided experimental access to this transitional regime.

Metallic SSL systems based on Ce or Yb moments have the potential for a more complex T=0 phase diagram~\cite{vojta2008,lacroix2009,coleman2010,aronson2010,bernhard2011}. In the absence of frustration,  both the Kondo temperature T$_{K}$ and the Ne\'{e}l temperature T$_{N}$ in systems of this sort can be tuned by pressure, magnetic fields, or doping, suppressing AF order among well-localized f-electrons to T=0 at a quantum critical point (QCP). The f-electrons may delocalize at or near this QCP, from which a strongly correlated paramagnetic phase with delocalized f-electrons emerges~\cite{stewart2001,vonlohneysen2007,gegenwart2008}. A crucial ingredient of dimer formation in the SSL is a doublet ground state, and crystal fields can produce such a pseudo-spin S=1/2 ground state in several Ce and Yb based heavy fermion (HF) compounds based on the SSL. Complex magnetic order is found in Ce$_{2}$Pd$_{2}$Sn, where novel low temperature properties arise from ferromagnetic (FM) dimers with the S=1 ground state ~\cite{sereni2009}. Nonordering Yb$_{2}$Pd$_{2}$Sn can be driven AF via pressure~\cite{bauer2008} and In doping~\cite{kikuchi2009}, but the T$_{K}$=17\emph{K} of Yb$_{2}$Pd$_{2}$Sn remains large throughout. No evidence for dimer formation, such as a singlet-triplet gap, is found and instead the magnetic susceptibility $\chi$ becomes constant as T$\rightarrow$0, indicating that Kondo physics dominates in Yb$_{2}$Pd$_{2}$Sn~\cite{bernhard2011}. In contrast, the Yb moments in \ypp~ remain fully localized, ordering at T$_{N}$=2.07 \emph{K}, with no indication of Kondo physics ~\cite{kim2008, aronson2010}.

We argue here that \ypp~ is a SSL system where frustration dominates over Kondo physics.  Measurements of the magnetic susceptibility $\chi$(T) find that \ypp~ is a quasi-two dimensional system where magnetic interactions are highly frustrated.  The dimer formation characteristic of the SSL is evidenced in \ypp~ by a broad maximum in $\chi$(T), suggesting that AF order in \ypp~ emerges from a paramagnetic dimer fluid with a singlet-triplet gap $\Delta$. Magnetic fields suppress AF order more quickly than the spin gap $\Delta$, indicating that the AF phase can only be entered from a spin liquid at higher temperatures and fields. Quantized magnetization steps are a signature of other SSL systems, such as \scbo~ and the RB$_{4}$ compounds, and they are observed as well within the AF  phase of \ypp. As a SSL system, \ypp~ exemplifies a regime near AF instability that has not previously been experimentally accessible. As a HF, \ypp~ is one of the first systems where the interplay of geometrical frustration and quantum criticality can be investigated.

All experiments were performed on single crystals of \ypp~ that were prepared from Pb flux~\cite{kim2008}. The electrical resistivity $\rho$ of \ypp~ is metallic and approaches a residual value $\rho_{0}$=1.5 $\mu\Omega$-cm, attesting to low levels of crystalline disorder~\cite{kim2008}. Measurements of the dc magnetization M were conducted at fixed fields ranging from 0.1 - 4 \emph{T} in a Quantum Designs Magnetic Properties Measurement System (MPMS) for temperatures from 1.8 \emph{K}-300 \emph{K}, while a Hall sensor magnetometer was used for temperatures from 0.06 \emph{K}-4 \emph{K}~\cite{wu2011}.

The dc magnetic susceptibility $\chi$=M/B (Fig.~1) reveals both quasi-two dimensionality and strong frustration in \ypp, where the Yb$^{3+}$ moments lie on the SSL (inset, Fig.~1a).
\begin{figure}
\includegraphics[scale=0.42]{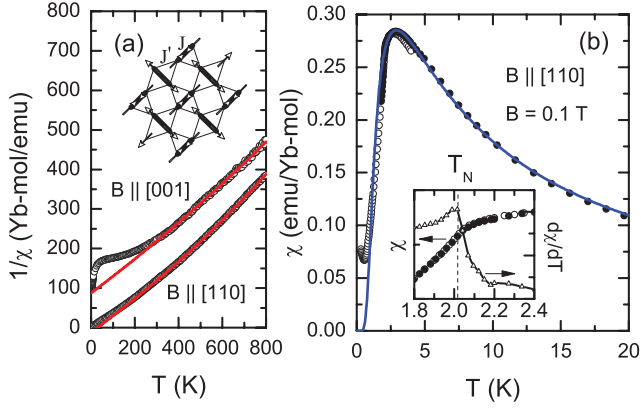}
\caption[]{(Color online) (a) Temperature dependencies of 1/$\chi$ for fixed fields B=2 \emph{T}(T$\geq$ 300 \emph{K}) and B=0.1 \emph{T} (T$\leq$ 300 \emph{K}) along [001] and [110]. Solid red lines are fits to Curie-Weiss expressions. Inset: the Shastry-Sutherland lattice has interdimer $J^{\prime}$ and intradimer \emph{J} couplings as indicated, with moments directed along the dimer bonds. (b) $\chi$(T)=M/B for B=0.1 \emph{T} (MPMS: $\bullet$, Hall sensor: $\bigcirc$). Solid line is fit to dimer expression (see text).  Inset: expanded view of region near T$_{N}$=2.02 \emph{K}. (vertical dashed line). $\chi$:$\bullet$, $\bigcirc$; d$\chi$/dT: $\vartriangle$. Solid line is a guide for the eye.}
\end{figure}
For 300 K$\leq$T$\leq$800 K, $\chi$ is well described by a Curie-Weiss temperature dependence $\chi$(T)=$\chi_{0}$+C/(T-$\theta$), where the fluctuating Yb moments are close to the 4.54 $\mu_{B}$/Yb$^{3+}$ Hund's rule value (Fig. 1a). Weiss temperatures  $\theta_{110}$=28 K (B$\|$[110]) and $\theta_{001}$=-217 K (B$\|$[001]), indicate weak FM correlations within the [110] SSL plane, but  stronger AF coupling between the SSL planes. The Yb moments are likely directed along the [110] and equivalent easy directions (Fig.~1a, inset)~\cite{kim2008}.  A slope discontinuity in $\chi$ and the accompanying  maximum in d$\chi$/dT marks the onset of AF order at T$_{N}$=2.02 \emph{K}, slightly below the T$_{N}$=2.07 \emph{K} that is found in specific heat measurements ~\cite{kim2008}(Fig.~1b, inset). AF order occurs in \ypp~ at a Ne\`{e}l temperature T$_{N}$ that is much smaller than the mean field values indicated by the Weiss temperatures, a hallmark of frustration~\cite{ramirez1994}. The in-plane frustration figure of merit f=$\theta_{110}$/T$_{N}$=14 and the interplanar f=$\theta_{001}$/T$_{N}$=105 reveal a profound interplanar frustration in \ypp~, indicating as does the large magnetic anisotropy $\chi_{110}$/$\chi_{001}$=30 (T=T$_{N}$), that the individual SSL planes remain magnetically  uncorrelated at the lowest temperatures.

A broad peak is observed in $\chi$(T) for B$\|$[110] (Fig.~1b), indicating that the ground state of \ypp~ is nonmagnetic. The magnetic susceptibility $\chi$(T) (B$\|$[110]) is well described using the mean field expression $\chi$(T)=$\chi_{D}$/(1-2n$J^{\prime}\chi_{D}$), where $J^{\prime}$ is the interdimer coupling, n the number of near neighbors, and $\chi_{D}$ is the susceptibility of a single dimer. Both of the Yb moments contribute two states, and coupling these moments into a dimer produces a singlet ground state and a triplet excited state, separated for B=0 by an energy $\Delta$= -2\emph{J}. $\chi_{D}$ is readily calculated from this energy level scheme~\cite{sasago1995}, taking N to be the number of dimers, k$_{B}$ the Boltzmann constant, $\mu_{B}$ the Bohr magneton, and g the Land\'{e} g-factor: \mbox{$\chi_{D}= (2N(g\mu_{B})^2/k_{B}T)[exp(-\Delta/k_{B}T)+3]$}. Although there is a small upturn in $\chi$(T) at the lowest temperatures, perhaps indicating that a few Yb moments or even stray impurity moments do not participate in the magnetic dimers, the fit (Fig.~1b) provides an excellent account of the measured B=0.1 \emph{T} susceptibility $\chi$(T) both above and below T$_{N}$ when $\Delta$=4.3$\pm$0.04 \emph{K}, \emph{J}=--2.3$\pm$0.01 \emph{K}, $J^{\prime}$=-1.95$\pm$0.03 \emph{K}, and g=5.43$\pm$0.02, the last consistent with observations in other systems where Yb$^{3+}$ is in a tetragonal crystal field~\cite{reynolds1972}.

 Magnetic fields affect both T$_{N}$ and $\Delta$, fundamentally changing the balance of phases present for \ypp~ at B=0.
 \begin{figure}
\includegraphics[scale=0.4]{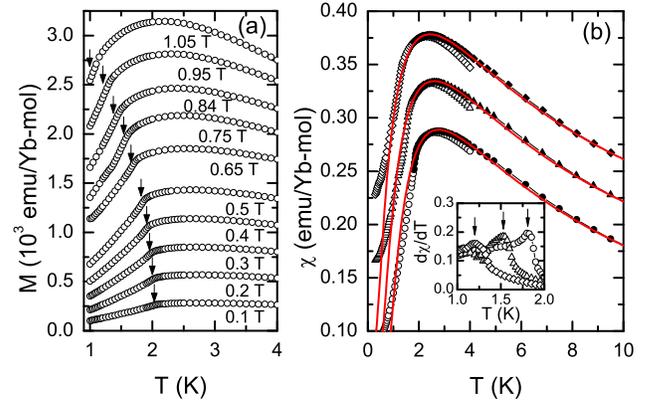}
\caption[]{(Color online)(a) Magnetization M for different fields B (indicated). Arrows mark T$_{N}$, taken from the peak in d(M/B)/dT. (b) $\chi$(T)=M/B for different values of B$\|$[110] ($\bigcirc$: 0.5 \emph{T}, $\bigtriangleup$: 0.75 \emph{T}, $\blacklozenge$: 0.95 \emph{T}). Solid red lines are mean field expression for $\chi$ (see text). Inset: arrows indicate peaks in d$\chi$/dT at T$_{N}$, obtained for B=0.5 \emph{T}, 0.75 \emph{T}, and 0.95 \emph(T) (right to left).}
\end{figure}
 Increasing magnetic fields B$\|$[110] shift both the slope discontinuity in M(T$_{N}$) (Fig.~2a) and its associated peak in d$\chi$/dT (Fig. 2b, inset), as well as the broad maximum in $\chi$(T)(Fig.~2b) to lower temperatures. T$_{N}$(B) is taken from the maximum in d$\chi$/dT (inset, Fig. 2b), and the values of T$_{N}$ determined for each field B are shown in Fig.~3a.  T$_{N}$ vanishes for  B$_{QCP}$=1.25$\pm$0.01 \emph{T}, following \mbox{T$_{N}\sim$ (B$_{QCP}$-B)$^{\nu}$} with the XY class exponent $\nu$=0.46$\pm$0.03 ~\cite{kawashima2004}. This behavior resembles that of HFs like YbRh$_{2}$Si$_{2}$~\cite{gegenwart2007} and CeRhIn$_{5}$~\cite{park2006} near their AF-QCPs. In contrast, the  Bose Einstein Condensation (BEC) exponent $\nu$=2/3 is found in quantum magnets like  BaCuSi$_{2}$O$_{6}$ \cite{sebastian2006} and TlCuCl$_{3}$ \cite{yamada2008}, where magnetic fields induce T=0  AF order  by driving  $\Delta\rightarrow$0, via the Zeeman splitting of excited triplet states~\cite{giamarchi2008}.

$\Delta$ and T$_{N}$ vanish at different fields in \ypp. The analysis of the B=0 $\chi$(T) can be generalized for B$\neq$0, using the energy level scheme depicted in Fig.~3a (inset).  Each dimer has a singlet ground state \mbox{(E$_{0}$=3/2 \emph{J})}, and three excited states with energies \mbox{E$_{1}$=-1/2\emph{J}-g$\mu_{B}$B},\mbox{E$_{2}$=-1/2\emph{J}}, and \mbox{E$_{3}$=-1/2\emph{J}+g$\mu_{B}$B}. The dimer magnetization M$_{d}$ is derived from the partition function of these four states, yielding the expression:

\begin{eqnarray*}
M_{d}=\frac{2g\mu_{B}\sinh\left(g\mu_{B}B/k_{B}T\right)}{1+\exp\left(-2J/k_{B}T\right)+2\cosh\left(g\mu_{B}B/k_{B}T\right)}
\end{eqnarray*}

The susceptibility $\chi$ of N interacting dimers, each with n neighbors, is given  in turn by the mean field expression \mbox{$\chi$(B,T)= N$\chi_{d}$/(1-2$J^{\prime}$n$\chi_{d}$)}, where the dimer susceptibility \mbox{$\chi_{d}$=dM$_{d}$/dB}. $\chi$(T) is calculated for each of the fields B represented in Fig.~2a, using the B=0.1 \emph{T} values of $J^{\prime}$=-1.95 \emph{K}, \emph{J}=-2.34 \emph{K}, and  g=5.43. The resulting expressions agree well with the measured $\chi$(T), shown in Fig.~2b.  The Zeeman splitting derived from this analysis gives \mbox{$\Delta$(B)= E$_{1}$-E$_{0}$ = -2 \emph{J}-g$\mu_{B}$B}, where$\Delta$ drops linearly from its B=0.1 \emph{T} value of 4.3 \emph{K} to zero for B$_{\Delta}$=1.5 \emph{T} (Fig.~3a, inset). We deduce that $\Delta$(B=0)=4.6 \emph{K}, by extrapolating the B=0.1 \emph{T} value $\Delta$=4.3 \emph{K} to B=0. The relative magnitudes of  \emph{J}=-2.3$\pm$0.01 \emph{K} and $J^{\prime}$= -1.95$\pm$0.03 \emph{K} extracted from the B=0.1 \emph{T} fit give the ratio $J^{\prime}/J$=0.85, a value that is larger than the critical value ($J^{\prime}/J$)$_{C}$= 0.6 -0.7, placing \ypp~ within the expected AF regime of the S=1/2 SSL~\cite{shastry1983}.
\begin{figure}
\includegraphics[scale=0.4]{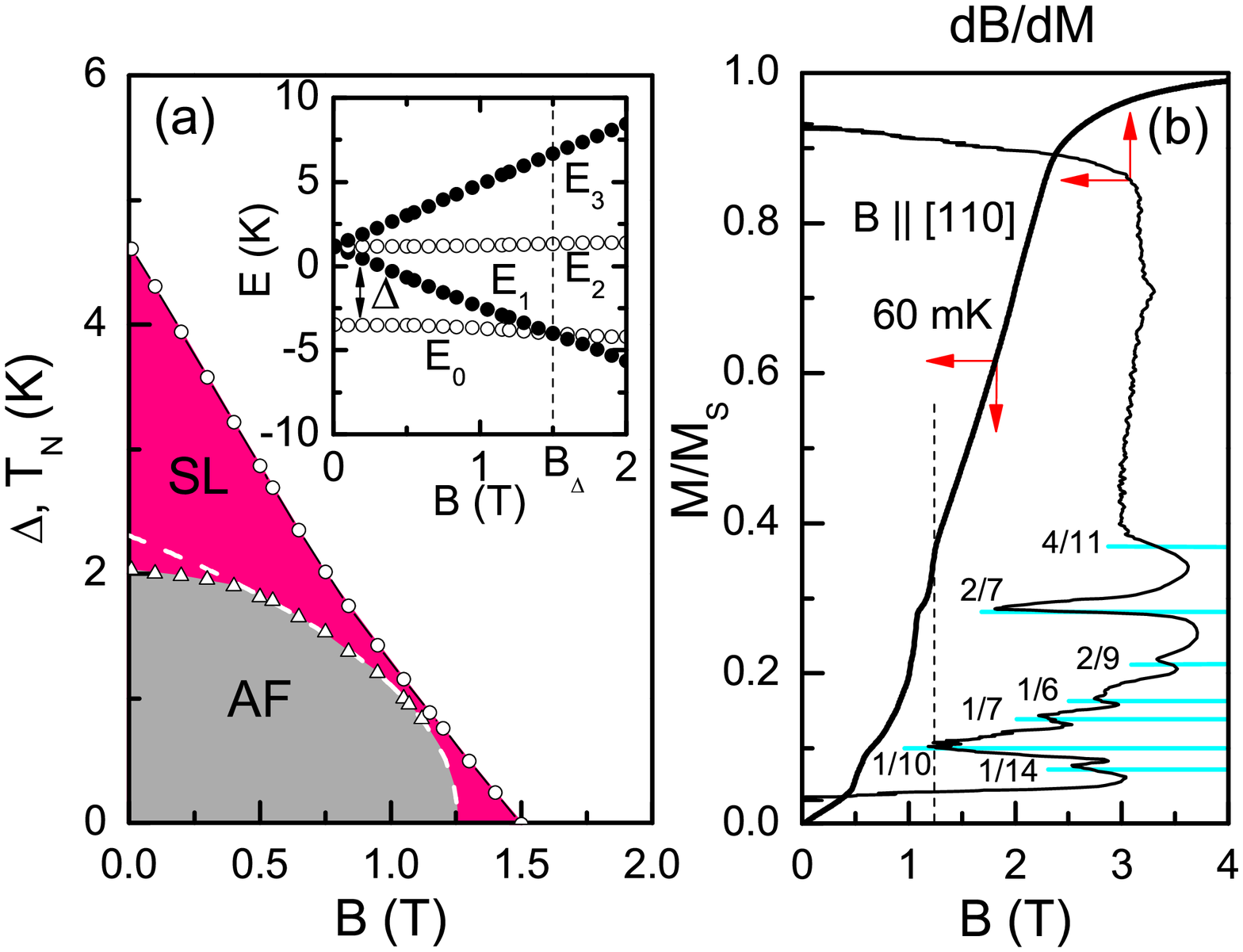}
\caption[]{(Color online) (a) The field dependencies of the Ne\'{e}l temperature T$_{N}$($\vartriangle$) and the singlet-triplet dimer gap $\Delta$ ($\bigcirc$). White dashed line is fit to \mbox{T$_{N}\simeq$(B$_{QCP}$-B)$^{0.46}$}. Black solid line is \mbox{$\Delta$=$\Delta_{0}$-g$\mu_{B}$B}, with $\Delta_{0}$=4.6 \emph{K} and g=5.43 obtained from fit (see text). Inset: Zeeman splitting of the excited dimer triplet states  with energies E$_{1}$, E$_{2}$, and E$_{3}$ (see text) leads to the vanishing of the singlet-triplet gap $\Delta$(B)=E$_{1}$-E$_{0}$ for B$_{\Delta}$=1.5 \emph{T} (vertical dashed line). (b) M(B), normalized to M$_{S}$=M(4 \emph{T}) for T=0.06 \emph{K}. Plateaux in M/M$_{S}$ (left axis) correspond to peaks in the inverse susceptibility dB/dM (right axis), with quantized values as indicated. Vertical dotted line indicates AF-QCP B$_{QCP}$=1.25 \emph{T}.}
\end{figure}

The phase diagram that is formed by comparing k$_{B}$T$_{N}$(B) and the energy scale $\Delta$(B)/k$_{B}$ (Fig.~3a) indicates that for B$_{QCP}\leq$B$\leq$B$_{\Delta}$, there is a nonzero singlet triplet gap $\Delta$, but no AF order. This regime can be considered a valence bond solid, where the ground state is a nonmagnetic singlet. The disappearance of $\Delta$ for B=B$_{\Delta}$ indicates that the singlet and triplet dimer states have become degenerate. In dimer systems like TlCuCl$_{3}$~\cite{ruegg2003} and BaCuSi$_{2}$O$_{6}$~\cite{sebastian2006}, this gapless and magnetic state is unstable to AF order, and T$_{N}$ increases as field increases the population of dimer triplets,  analogous to BEC. In \ypp~, the B=0 AF phase has already vanished when $\Delta\rightarrow$0,  although it is possible that re-entrant AF order or another collective state may result for B$\geq$B$_{\Delta}$~\cite{yoshida2005}.

Perhaps the most striking signature of the SSL is the observation of quantized steps in M(B), present either in the field-induced AF phase in \scbo~\cite{kageyama1999,sebastian2008}, or in the AF phase that is present for B=0 in the RB$_{4}$~\cite{siemensmeyer2008,yoshii2008}.  We note that they may alternatively be metamagnetic transitions resulting from the interplay of exchange and magnetocrystalline anisotropy that is found in unfrustrated systems like CeSb~\cite{rossat1983}.  \ypp~ is like the other SSL systems, as a sequence of magnetization plateaux are evident as broadened steps in M(B) or sharp peaks in dM/dB, measured at T=0.06 \emph{K} (Fig.~3b). Increasing and decreasing field sweeps are hysteretic, indicating that \ypp~ approaches full saturation with M$\rightarrow$M$_{S}$ via a series of intermediate phases that are separated by first order transitions, each with increasing fractions of dimer triplets aligned with the external field. Fig.~3b shows that the M(B) plateaux are only observed in \ypp~ in the AF state with B$\leq$B$_{QCP}$. Unlike \scbo, TlCuCl$_{3}$, and BaCuSi$_{2}$O$_{6}$, where very large fields are required to approach saturation, in \ypp~ M/M$_{S}\rightarrow$1 for B$\simeq$4 \emph{T}, so it is straightforward to observe the entire magnetization process.

Our experiments on \ypp~ provide new insight into AF order on the SSL. \ypp~ is a conventional paramagnet when k$_{B}$T$\gg J, J^{\prime}$, but an increasing number of Yb moments form long-lived dimers as k$_{B}$T decreases towards $\Delta$=4.6 \emph{K}. The stabilization of AF order requires a substantial occupancy of the excited moment-bearing triplet state, which is only possible when k$_{B}$T$_{N}$ is not much smaller than $\Delta$. \ypp~ is the only known SSL system where this condition is met, and the apparent persistence of the singlet-triplet gap into the AF state suggests that AF order involves locking strongly bonded dimers together via weaker interdimer bonds. The phase diagram in Fig.~3a indicates that increasing either temperature or magnetic field breaks these fragile interdimer bonds, and \ypp~ reverts to a  liquid of uncoordinated dimers.

The unique characteristics of \ypp~ are highlighted by comparing its properties to other SSL systems (Table~1). The  magnitudes of the moments, as well as the Weiss temperatures for fields in the SSL plane $\theta_{ab}$ are similar for the RB$_{4}$ and \ypp, and consequently \ypp~ and the RB$_{4}$ might be expected to order at similar temperatures. However, AF order is only found in \ypp~ when T$_{N}\ll\theta_{ab}$, resulting, in part, from the quasi-two dimensional character of $\chi$ in \ypp, absent in the other SSL compounds. This suppression of k$_{B}$T$_{N}$ to a value that is comparable in magnitude to $J^{\prime}$ and \emph{J} makes dimer formation an integral feature of AF \ypp, making \ypp~ the AF counterpart of the SL \scbo. As such, it is the only SSL system where the interplay of dimer formation and long-ranged AF order can be studied.

\begin{table}
\caption{\label{tab:table1}A comparison of the Ne\'{e}l temperature T$_{N}$, in-plane Weiss temperature $\theta_{ab}$, frustration figure of merit f=$\theta_{ab}$/T$_{N}$, interdimer exchange $J^{\prime}$, intradimer exchange \emph{J}, $J^{\prime}/J$, and susceptibility anisotropy $\%$= $\chi_{ab}$/$\chi_{c}$, evaluated at T$_{N}$ in different SSL systems (2 \emph{K} for \scbo, 250 \emph{K} and 30 \emph{K}, respectively, for TmB$_{4}$). T$_{N}$, $\theta_{ab}$ $J^{\prime}$, and \emph{J} are all given in units of \emph{K}. }
\begin{ruledtabular}
\begin{tabular}{|c|c|c|c|c|c|c|c|c|}
&T$_{N}$&$\theta_{ab}$&f&\emph{J}&$J^{\prime}$&$J^{\prime}/J$&$\%$&REF.\\
\hline Yb$_{2}$Pt$_{2}$Pb           &2.02 &28 &14    &2.3    &1.9    &0.83   &30     &This work\\
\hline GdB$_{4}$                    &42    &-68  &1.6  &8.9    &0.68   &0.076   &1.05   &\cite{fisk1981,fernandez2005,kikkawa2007}\\
\hline TmB$_{4}$                    &10  &-63  &6.3   &0.85   &0.3    &0.36   &1.5,20    &\cite{matas2010}\\
\hline TbB$_{4}$                    &44    &-27   &0.6 &1.55   &0.33   &0.21   &0.88   &\cite{fisk1981,rhyee2007,yoshii2008}\\
\hline SrCu$_{2}$(BO$_{3}$)$_{2}$   &--    &-103    &--   &85      &54    &0.64   &1.28  &\cite{kageyama1999b,miyahara2003}\\
\end{tabular}
\end{ruledtabular}
\end{table}

The B=0 ground state for \ypp~ is distinct among both HF and SSL compounds, with AF order developing from a liquid of dimers. The low Ne\'{e}l temperature, the persistence of the singlet-triplet dimer gap $\Delta$ in the AF state,  and the suppression of AF order in a small magnetic field  all place \ypp~ very close to the AF-SL transition, a regime of the SSL that was previously only addressed theoretically. While there are other HFs that form on geometrically frustrated lattices~\cite{lacroix2009}, in these cases it is generally found that long ranged interactions such as the Rudermann-Kittel-Kasuya-Yosida (RKKY) interaction replace the competing short ranged interactions that lead to frustration effects in insulating systems.  Given that \ypp~ is an excellent metal with substantial Yb moments, it is noteworthy that we observe the  singlet-triplet gap, the dimer spin liquid, and the magnetization plateaux, all signatures of the SSL that were previously only observed in insulating \scbo. It is at present unknown whether the HF character of \ypp~ will result in the same breakdown in normal metallic behavior and the stabilization of unconventional ordered phases that are found near AF quantum critical points in unfrustrated HF compounds, or if HFs with geometrical frustration have inherently different properties. \ypp~ is one of a very small number of known compounds where these intriguing questions can be experimentally explored.

The authors thank J. Sereni for enlightening discussions. This work was supported by National Science Foundation grant NSF-DMR-0907457.

\end{document}